\documentclass[conference]{IEEEtran}
\ifCLASSINFOpdf
   \usepackage[pdftex]{graphicx}
   \graphicspath{{../pdf/}{../jpeg/}}
\else
\fi
\usepackage{amsmath,amssymb}
\usepackage{algorithm}
\usepackage[noend]{algpseudocode}

\usepackage{booktabs}
\usepackage{array}
\usepackage{fancyhdr}
\usepackage[pscoord]{eso-pic}

\begin{document}
\title{Using Reinforcement Learning for Demand Response of Domestic Hot Water Buffers: a Real-Life Demonstration}



%
\author{\IEEEauthorblockN{Oscar De Somer\IEEEauthorrefmark{1}\IEEEauthorrefmark{2},
Ana Soares\IEEEauthorrefmark{1}\IEEEauthorrefmark{2},
Tristan Kuijpers\IEEEauthorrefmark{3}, Koen Vossen\IEEEauthorrefmark{3},
Koen Vanthournout\IEEEauthorrefmark{4} and
Fred Spiessens\IEEEauthorrefmark{1}\IEEEauthorrefmark{2}}
\IEEEauthorblockA{\IEEEauthorrefmark{1}Flemish Institute for Technological Research (VITO), Boeretang 200, B-2400 Mol, Belgium}
\IEEEauthorblockA{\IEEEauthorrefmark{2}EnergyVille, Thor Park Poort Genk 8130, 3600 Genk, Belgium}
\IEEEauthorblockA{\IEEEauthorrefmark{3}Enervalis,
Greenville Campus, Centrum Zuid 1111, Houthalen-Helchteren, Belgium}
\IEEEauthorblockA{\IEEEauthorrefmark{4}ENION, Kromstraat 16, 3740 Bilzen, Belgium}
}

\maketitle


\begin{abstract}
This paper demonstrates a data-driven control approach for demand response in real-life residential buildings. The objective is to optimally schedule the heating cycles of the Domestic Hot Water (DHW) buffer to maximize the self-consumption of the local photovoltaic (PV) production. A model-based reinforcement learning technique is used to tackle the underlying sequential decision-making problem. The proposed algorithm learns the stochastic occupant behavior, predicts the PV production and takes into account the dynamics of the system. A real-life experiment with six residential buildings is performed using this algorithm. The results show that the self-consumption of the PV production is significantly increased, compared to the default thermostat control.
\end{abstract}


\hfill Reinforcement Learning, Demand Response, Domestic Hot Water, Field Experiment
\hfill March 1, 2017

%
\IEEEpeerreviewmaketitle

\section{Introduction}
\thispagestyle{fancy}
\lhead{Submitted to IEEE ISGT Europe 2017}

Residential demand response (DR) has received considerable attention in the recent literature. It can enable consumers with flexible loads to adapt their consumption profile in response to an external signal. The increasing amount of installed PV has a significant impact on the existing power system. It accelerates transformer aging and may cause voltage problems on the distribution feeder \cite{Gonzalez2012}. Due to these problems, the tariffs for injecting energy in the grid are decreasing, and are even zero in some countries.

Residential DHW buffers offer the possibility to store thermal energy without impacting the user comfort. Since the heating system is used for a double purpose and an insulated buffer is inherently cheaper than electrochemical energy storage, the investment costs are lower compared to battery systems. Besides that, it does not induce as much energy losses as with grid connected battery systems \cite{fares2017the}.

Several research projects investigate the potential and implementation of residential demand response. Field tests in the residential sector have been conducted in multiple countries in order to assess the response to various input signals and the resultant flexibility \cite{Dupont2012, Kazmi2016, Leurs2016}.

This project comprises the control of thermostatically controlled loads (TCL), which is a stochastic sequential optimization problem. Model predictive control (MPC) and reinforcement learning (RL) are two candidate approaches to solve such problems. MPC was originally designed to exploit an explicitly formulated model of the process and solve in a receding horizon manner a series of open-loop deterministic optimal control problems. RL was designed to infer closed-loop policies for stochastic optimal control problems from a sample of trajectories gathered from interaction with the real system or from simulations \cite{Ernst2009}. RL has already been applied in several related test cases with promising results \cite{Leurs2016, ruelens2016TCL, ruelens2015residential}.

This paper describes the implementation and the results of a demand response application. In the context of the Rennovates project \cite{rennovates_webpage}, houses in social districts in the Netherlands are renovated and equipped with a smart heat pump and a PV installation. The objective is to optimally schedule the heating cycles of the DHW buffer to maximize the self-consumption of the local PV production. To do so, a model-based RL approach is employed. The main contribution of this paper is the \emph{real-life} implementation of a PV self-consumption algorithm using non-battery residential devices.

Section \ref{sec:setup} discusses the demonstration set-up. After that, section \ref{sec:mdp} presents the formal notation that is used in this work and its implementation in the real-life set-up. In section \ref{sec:control} the control strategy is described and section \ref{sec:results} discusses the results. Conclusions and future research directions are finally outlined in section \ref{sec:conclusion}.

\section{Set-Up}
\label{sec:setup}
In its initial configuration, the field experiment consists of six renovated houses, however, more houses will be added in a later phase. In each house, a smart heat pump is used for space heating and for heating the DHW buffer. The insulation of the houses and the number of PV panels are dimensioned such that the annual energy production exceeds its consumption. The following three subsections will give a concise description of the DHW buffer, the available sensors, and the controls.

\subsection{DHW Buffer}
The DHW buffer has a volume of 200 liters. The buffer remains properly stratified when water is tapped, but gets mixed when a heating cycle is started. Since this mixing creates an unpredictable temperature distribution, only complete charging cycles are allowed, i.e., the heating process is only stopped when the temperature distribution in the buffer is assumed to be uniform. This considerably simplifies the problem of estimating the energy content of the buffer.

\subsection{Sensors}
A collection of sensors provide input for the control algorithm. The most important for this use case are: a temperature sensor in the middle of the buffer, a flow meter to measure the volume of hot water tapped, and an electricity meter of the heat pump.

\subsection{Controls}
The charging cycles of the DHW buffer can be controlled by two control commands on the smart heat pump. As first, the target temperature of the buffer can be set ($T_{set}$). This means that when the temperature, measured by the sensor in the middle of the buffer, decreases below $T_{set}$, the heat pump starts charging to $T_{set}$. In order to ensure the user comfort limits, the minimum allowed target temperature is $T_{min} = 45^\circ C$, and maximum $T_{max} = 55^\circ C$. Secondly, a control command is available to force start the heating process. This control can be used when a significant portion of hot water is tapped but the cold water front is still below the temperature sensor in the middle of the buffer. In this case, the sensor does not measure the cold water yet, thus the heating would not start automatically.

\section{Markov Decision Processes}
\label{sec:mdp}
This section introduces the notation used in the remainder of this work. In RL, a problem is usually formulated as a Markov Decision Process (MDP) \cite{sutton1998reinforcement, Ernst2005}. An MDP is defined by its state space $X$, its action space $U$, and a transition function $f$:
\begin{equation}
\label{eq:transf}
x_{k+1} = f(x_k, u_k, w_k), \quad k=0,1, \dots, T-1,
\end{equation}
which describes the dynamics from $x_k$ to $x_{k+1}$, under control action $u_k \in U$, and subject to random disturbance $w_k \in W$. The number of control periods in one optimization horizon is represented by $T$. The disturbance $w_k$ is generated by a conditional probability distribution $p_w(\cdot|x)$. The transition from $k$ to $k+1$, is associated with a cost $c_k$:
\begin{equation}
c_k = \rho(x_k, u_k, w_k), \quad k=0,1, \dots, T-1,
\end{equation}
it corresponds to a cost for injecting energy in the grid. The aim is to find the control policy $h: X \rightarrow U$ that minimizes the expected sum of costs over the considered time period:
\begin{equation}
J^h(x_t) = \mathbb{E}(\sum_{k=t}^{T-1}\rho(x_k, h(x_k), w_k)).
\end{equation}

Typically, this policy is characterized by a state-action value function or Q-function:
\begin{equation}
\label{eq:Qfun}
Q^h(x, u) = \underset{w \sim p_w(\cdot|x)}{\mathbb{E}}[\rho(x, u, w) + J^h(f(x, u, w))],
\end{equation}
which estimates the expected cost when choosing action $u$ in state $x$, and following policy $h$ thereafter. Given a Q-function, the policy is calculated by choosing an action that minimizes the expected cost in a given state:
\begin{equation}
h(x) \in \underset{u \in U}{\arg\!\min} Q^h(x,u).
\end{equation}

The next four paragraphs describe the state, action, transition function, and cost function in detail.

\subsection{State Description}
\label{subsec:state_descr}
The relevant state space $X$ of a domestic water heater can be broken down in a time-related component, a controllable component and a non-controllable exogenous component \cite{ruelens2015residential, ruelens2016TCL}.

\subsubsection{Time feature}
The time-related component represents the discrete feature space that contains relevant timing information. Using this information the learning algorithm should be able to capture the dynamics of the system.

For residential consumers, the behavior comprises a daily and a weekly pattern. Therefore, the most relevant timing information can be captured by the hour of the day: $X_t^h = \left\{1, \dots, 24\right\}$ and the day in the week: $X_t^d = \left\{1, 2\right\}.$

Note that the day in the week is either a weekday ($x_t^d = 1$) or a weekend day ($x_t^d = 2$). By only differentiating between those two options, the main weekly pattern can emerge from a minimal number of costly data samples.

\subsubsection{Boiler representation}
The controllable component describes the state of the DHW boiler, which is influenced by control actions. The state of the boiler corresponds to the temperature distribution 	over the height of the buffer. However, as described in section \ref{sec:setup}, only one temperature sensor in the middle of the buffer is used.

The most important information derived from this temperature distribution is the thermal energy content of the boiler. A heuristic approach to estimate the energy content in a stratified buffer can be obtained with a 2-layer model. In this case, the boiler model has only two separate temperature layers. An upper layer that contains heated water, with a temperature set to be equal to the temperature measured by the sensor in the middle of the buffer: $T_{hot}$. The lower layer contains cold input water, with a temperature approximated by the feeding water temperature $T_{in}$.

When the boiler is uniformly charged, the boundary between the two layers is completely at the bottom. The boundary moves upward according to the volume of hot water tapped by the consumers since the last heating cycle $V_{tapped}$, as described in section \ref{sec:setup}.

Using this model the energy content of the buffer can be estimated as follows:
\begin{equation}
E = c_p(T_{hot}(V_{buffer}-V_{tapped}) -  T_{in}V_{tapped}),
\end{equation}
with $c_p$ the heat capacity of water and $V_{buffer}$ the volume of the DHW buffer. This model assumes that temperature distribution in the hot layer is uniform. Therefore, it is important to reset $V_{tapped}$ to zero regularly by fully recharging to ensure the model is correct. The field experiments show that this happens typically once a day.

$T_{in}$ is not directly measured by the system. This temperature slowly changes over time according to the outdoor temperature. An estimate of this temperature can be made by using the minimum temperature measured over the past month. The field experiments show that at least once a month more than 100 liters of water is tapped before the heat pump turns on. On such occasion the cold water front passes the temperature sensor in the middle of the buffer, which makes the input temperature directly measurable by this sensor.

The accuracy of this model decreases with increasing $V_{tapped}$. Since all the houses use the same set-up, the heuristic model was validated by measuring the real energy content in an identical lab set-up. Six temperature sensors were placed under the insulation material of the buffer. Those real energy measurements can then be used to fit a regression model that learns the 2-layer model error in function of $V_{tapped}$.

The remainder of this work will use state of charge ($SoC$), defined as:
\begin{equation}
SoC = (E - E_{min}) / (E_{max} - E_{min}),
\end{equation}
where $E_{max}$ is defined as the energy content of the buffer when it is uniformly charged at $T_{max}$:
\begin{equation}
E_{max} = c_p T_{max}V_{buffer},
\end{equation}
and $E_{min}$ is defined as the energy content of the buffer when half of it is equal to $T_{in}$ and the other half equal to $T_{min}$:
\begin{equation}
E_{min} = c_p(T_{in}V_{buffer}/2 + T_{min}V_{buffer}/2).
\end{equation}


\subsubsection{Exogenous information}
\label{subsub:exo_info}
The non-controllable component of the state space contains exogenous information of the system. In this case, the exogenous component only consists of a forecast of the PV production. It has no direct influence on the dynamics, but determines the cost function $\rho$.

A data driven approach is used to obtain a PV power forecast. It takes as input two data streams. The first is the historic PV power values, where the amount of historic data is a parameter. The second is the weather data forecast at the physical location of the site. This data stream comes from forecast.io \cite{forecastio}, an open source weather forecasting API, that contains weather features $p_{ex}$.

The forecaster is implemented to represent the function:
\begin{equation}
f: (p_{t-n}, \dots, p_{t-1}, p_{ex}) \rightarrow (p_t, \dots, p_{t+k}),
\end{equation}

where $n-1$ is the number of historic power values taken into account, and $k$ is the amount of future prediction. A deep neural network is trained on a set of samples of the form $(x, f(x))$ coming from the historic data (i.e. also taking historic weather data into account). To perform the forecast, all the required data is collected and used as input for the neural network. This gives the PV power prediction.

\subsection{Action}
As described in section \ref{sec:setup}, the target temperature of the boiler can be set and a heating cycle can be forced to start. The considered actions are to either delay the reheating by setting the target temperature to $T_{min}$, or  to start the heat pump and charge the buffer uniformly to a certain temperature. In the remainder of this work the action space will consist of three options, i.e., $u \in \left\{0, 1, 2\right\}$, with:
\begin{itemize}
\item $u=0$: Delay heating
\item $u=1$: Start heating now until $SoC$ corresponding to buffer with uniform temperature equal to $T_{min}$
\item $u=2$: Start heating now until $SoC = 1$
\end{itemize}

\subsection{Transition Function}
This work proposes a model-based reinforcement learning approach. That means that a model of the transition function is used to generate data as input for the control strategy described in section \ref{sec:control}. This model is learned using historical measurement data. It needs to describe the evolution in $SoC$ over time, contain the stochastic user behavior and incorporate the losses to the environment. On top of that, there is backup controller (BUC) present in the system that avoids comfort limits violations. This BUC should also be contained in the model.

\subsubsection{Tap water model}
\label{subsub:tap_water}
The user behavior is captured by constructing an approximation of the conditional probability distribution $p_{v}(\cdot|t)$, that can be used to generate samples $v_i$ of the water tapped at a given time step. It is $v_i$ that represents $w_k$ from (\ref{eq:transf}). This is done by binning the historical tap water consumption data points into bins corresponding to their time-related component of the state space. For the time feature discussed in section \ref{subsec:state_descr}, all the data points from the same hour in the day and same week day are gathered in the same bin. Given a certain timestamp, samples are generated by sampling randomly from the bin corresponding to this timestamp. If enough historical data is available, this generator will be a good representation of the underlying stochasticity. In this work, the last two months of consumption data are used for training the model.

A disadvantage of the discussed model is that it neglects the inter time step dependency of the tap water consumption. However, auto-correlation plots show only minor correlation with previous time steps. In future work it will be verified whether generating complete trajectories that take into account past consumption data has a positive influence on the results.

\subsubsection{Standing losses}
Besides the decrease in SoC by the water tapped from the buffer, there are standing losses to the environment. These standing losses are approximated by a regression function only dependent of the SoC, that is learned from historical data.

\subsubsection{Backup controller} As last, the transition function should contain the BUC, which corresponds to the automatically started heating cycles when the SoC decreases below zero. This is modeled by increasing the SoC back to the SoC corresponding to a buffer with a uniform temperature equal to $T_{min}$. Note that, the effect of taking action 1 is equal to action 0 when the $SoC$ is lower or equal to zero.

\subsection{Cost Function}
In this work the objective is to maximize the self-consumption of the PV production. Hence, the cost function is defined as: $\rho(x_k, u_k, w_k) = P_{inj} \Delta t$, where $P_{inj}$ is the average power injected in the grid during the time interval $\Delta t$, which is defined as the PV production minus the PV production covered by the heat pump consumption corresponding to the chosen action. A forecast of the PV production is available as described in section \ref{subsub:exo_info}. Obviously, this cost function can easily be adapted to offer more flexible demand response options. Note that non-controllable loads are ignored here since they are very hard to predict for an individual house.

\begin{figure}
\centering
\includegraphics[width=0.48\textwidth]{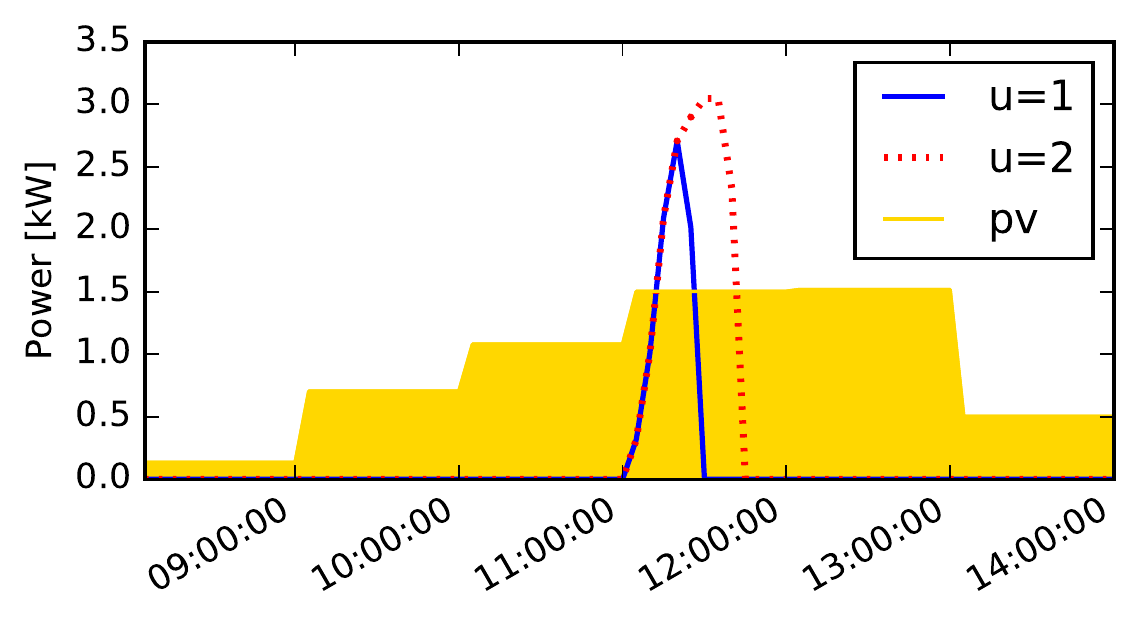}
\caption{Example PV production forecast and power profile heat pump for different actions and starting from a SoC of 0.25.
}
\label{fig:hp_prof}
\end{figure}

The power consumption of the heat pump depends on the chosen action and the $SoC$ of the buffer. However, the power profile can not be modulated, only the duration of the consumption can be elongated. This means that given the SoC of the boiler and the target temperature, the power profile can be deduced. Fig. \ref{fig:hp_prof} shows an example of the power profiles for two actions.

\section{Control Strategy}
\label{sec:control}
The central idea behind batch reinforcement learning is to estimate the Q-function (\ref{eq:Qfun}) based on past observations. Those observations are represented by 4-tuples $(x, u, x', c)$, containing the state, action, next state and corresponding cost.

In this work, first a model is learned to generate observations and then this model is used to build a training set to learn the policy. This approach reduces the number of interactions with the system. Note that neither the dynamics of the system nor the cost function are given in an analytical form. The optimal behavior in this strongly stochastic environment should be learned by interacting with the model \cite{ernst2005approximate}. In order to simplify the estimation of the Q-function, a separate approximation $\hat{Q}_N$ is made at every time step. Another simplification is made by iterating \emph{backwards} over the time steps, i.e., start at the end of the optimization horizon, set the Q-function equal to zero there and run backward in time. Using this strategy the Q-function contains all the information about the future costs after one sweep over the optimization horizon.

\begin{algorithm}[t]
\caption{Model based fitted Q-iteration \cite{ernst2005approximate}}\label{alg:rennoQ}
\begin{algorithmic}
\Require Grid $X \times U$ over state-action space, transition function $f$ and cost function $\rho$
\State $\hat{Q}_{T+1} \gets 0$
\For{$N=T,\cdots ,1$}
\State// Build training set $\mathcal{TS} = \left\{ (i^l, o^l) \right\}_{l=1}^{|\mathcal{F}|}$:
\State $l \gets 0$ 
\For{$\forall(x,u) \in X \times U$} 
\State $l \gets l + 1$
\State $i^l \gets (x, u)$
\State Get $L$ tap water samples $\left\{v^i\right\}_{i=1}^L$ for time step $k$
\State $o^l \gets \dfrac{\sum\limits_{i=1}^{L}\left[\rho(x, u, v^i) + \underset{u \in U}{\min} \hat{Q}_{N+1}\left( f(x, u, v^i),u\right)\right]}{L}$
\EndFor
\State \textbf{end for}
\State Use regression algorithm to fit $\hat{Q}_N$ from $\mathcal{TS}$
\EndFor
\State \textbf{end for}
\end{algorithmic}
\end{algorithm}

Algorithm \ref{alg:rennoQ} describes the calculation of the sequence of $\hat{Q}_N$-functions. As described in section \ref{subsub:tap_water}, only a timestamp is needed to generate tap water samples. For each \hyphenation{time-stamp} in the considered period, a grid is created over the state-action space. For each grid point, $L$ samples are generated from the tap water model. Those samples are used to calculate one output $o^l$ of the training set, with the grid point as input $i^l$. As can be seen, the random component in the cost function $\rho$ and the transition function $f$ is replaced by a sample from the tap water model $v \sim p_v(\cdot|t)$. A regression algorithm is used to generalize this information to any unseen state-action pair, i.e., a function approximation is used to fit this training set in order to get an approximation $\hat{Q}_N$ of $Q_N$ over the whole state-action space. In this case, an ensemble of extremely randomized trees is used as supervised learning algorithm \cite{geurts2006extremely}.

\section{Results}
\label{sec:results}
This section presents the results of the proposed algorithm in a real-life demonstration. The experiment considers six houses from the same district over a period of four months. A time step of one hour is used to calculate the $\hat{Q}$-functions. The state space is partitioned in 25 equidistant grid points and at every grid point 200 samples of the tap water model are taken. Every hour the policy, for a receding horizon of 24 hours, is retrained with the latest data. Every five minutes the best action is chosen.

The next two subsections investigate the learned control policy and some typical active control behavior will be discussed. Besides that, the performance of the learning algorithm is compared with the default scenario using static thermostat control.

\subsection{Control Policy}
\begin{figure}
\centering
\includegraphics[width=0.48\textwidth]{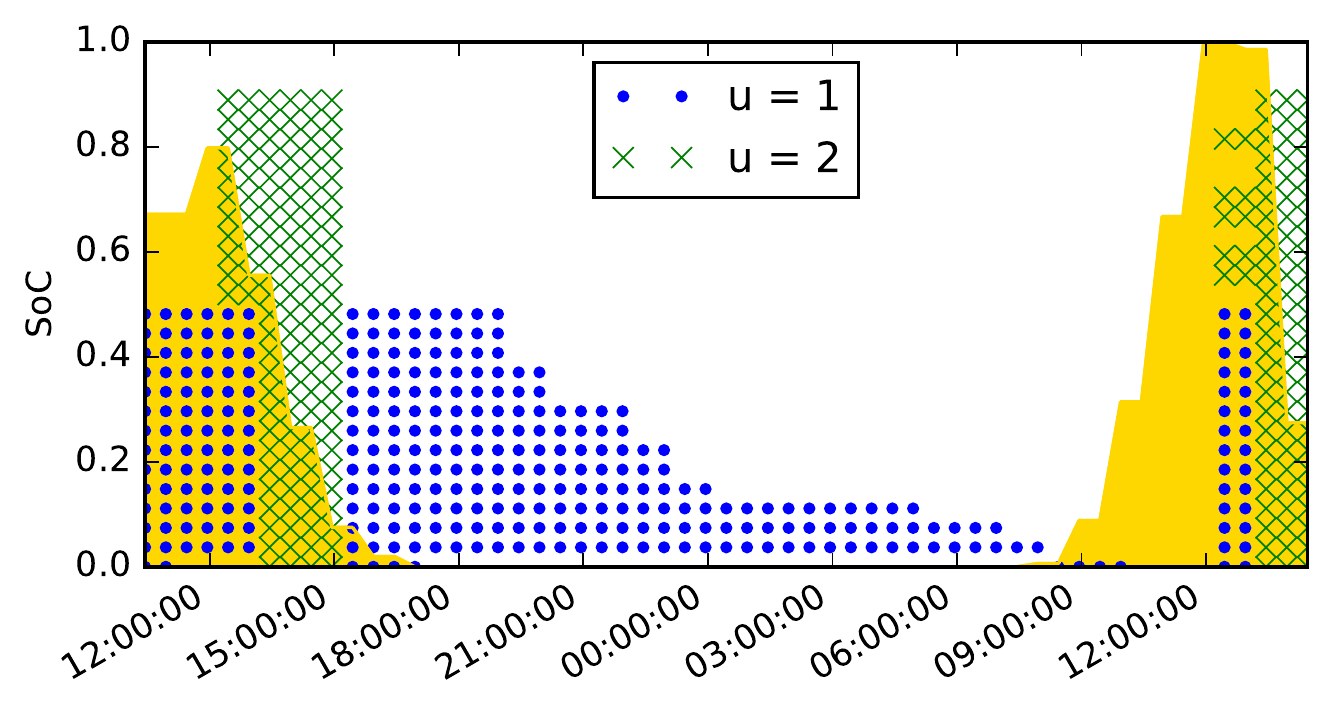}
\caption{Control policy over a period of 28 hours. The yellow background area shows the normalized predicted PV production. The dotted area depicts the state space where the policy recommends to charge the boiler to the minimum temperature (u = 1). The crossed area corresponds to the state space where the boiler should start charging to its maximum temperature (u = 2). On the state space indicated by the white area heating should be delayed (u = 0).}
\label{fig:policy}
\end{figure}

\begin{figure}
\centering
\includegraphics[width=0.48\textwidth]{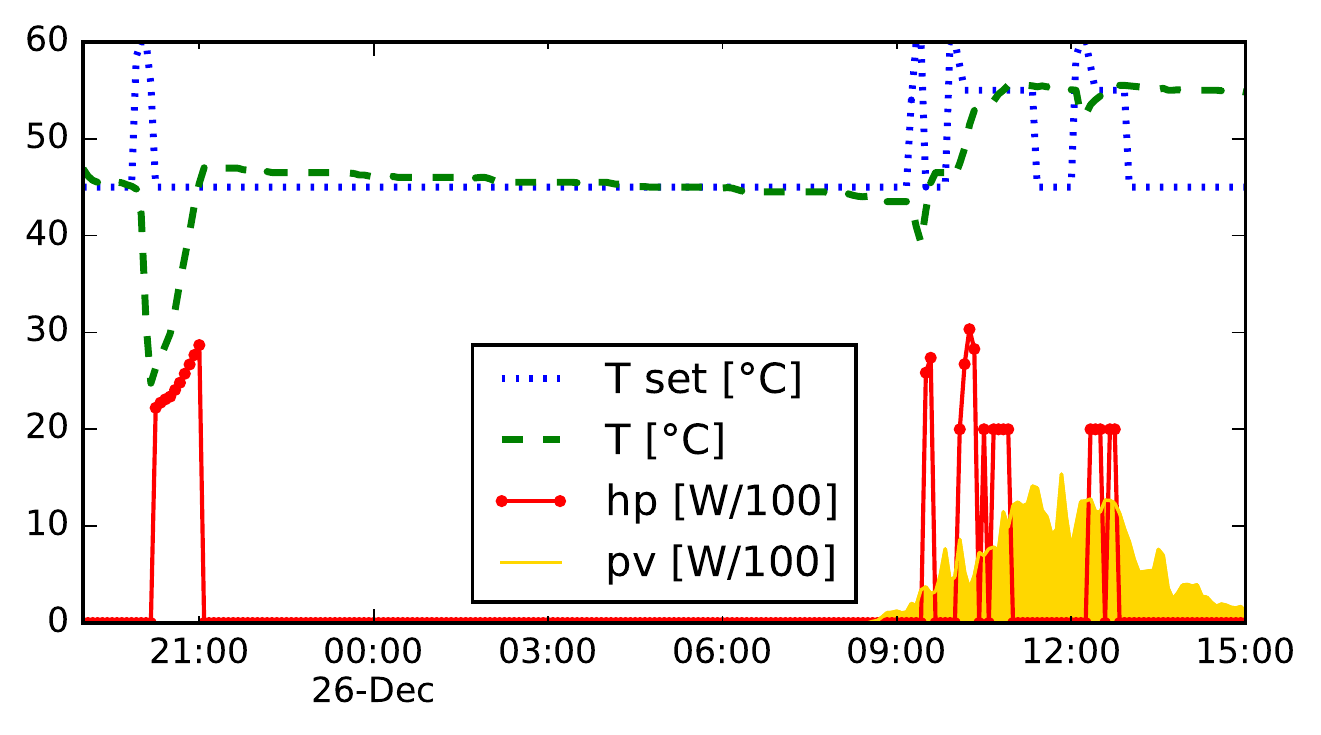}
\caption{Example of real trajectory using learned control policy. The dotted and dashed lines represent, respectively, the set point temperature and the temperature in the middle of the buffer. The yellow background area represents the measured PV production. The solid line represents the heat pump consumption for charging the boiler.}
\label{fig:example_traject}
\end{figure}

Fig. \ref{fig:policy} shows an example of the learned control policy over a period of 28 hours. It shows the best action, according to the policy, over the state space. It can be seen that the policy tries to avoid charging in the period close before the highest PV production. Depending on the state of charge and the time, it charges during the night to the minimum temperature to avoid a charging cycle induced by the backup controller right before noon. When the PV production comes close to its maximum it starts heating the boiler. Preferably, first to the minimum temperature (u = 1) and then to the maximum temperature (u = 2). When the PV production starts declining, it will charge to its maximum temperature if possible (u = 2). An example of a real trajectory on a winter day is shown in Fig. \ref{fig:example_traject}.

\begin{figure*}
\includegraphics[width=0.91\textwidth
]{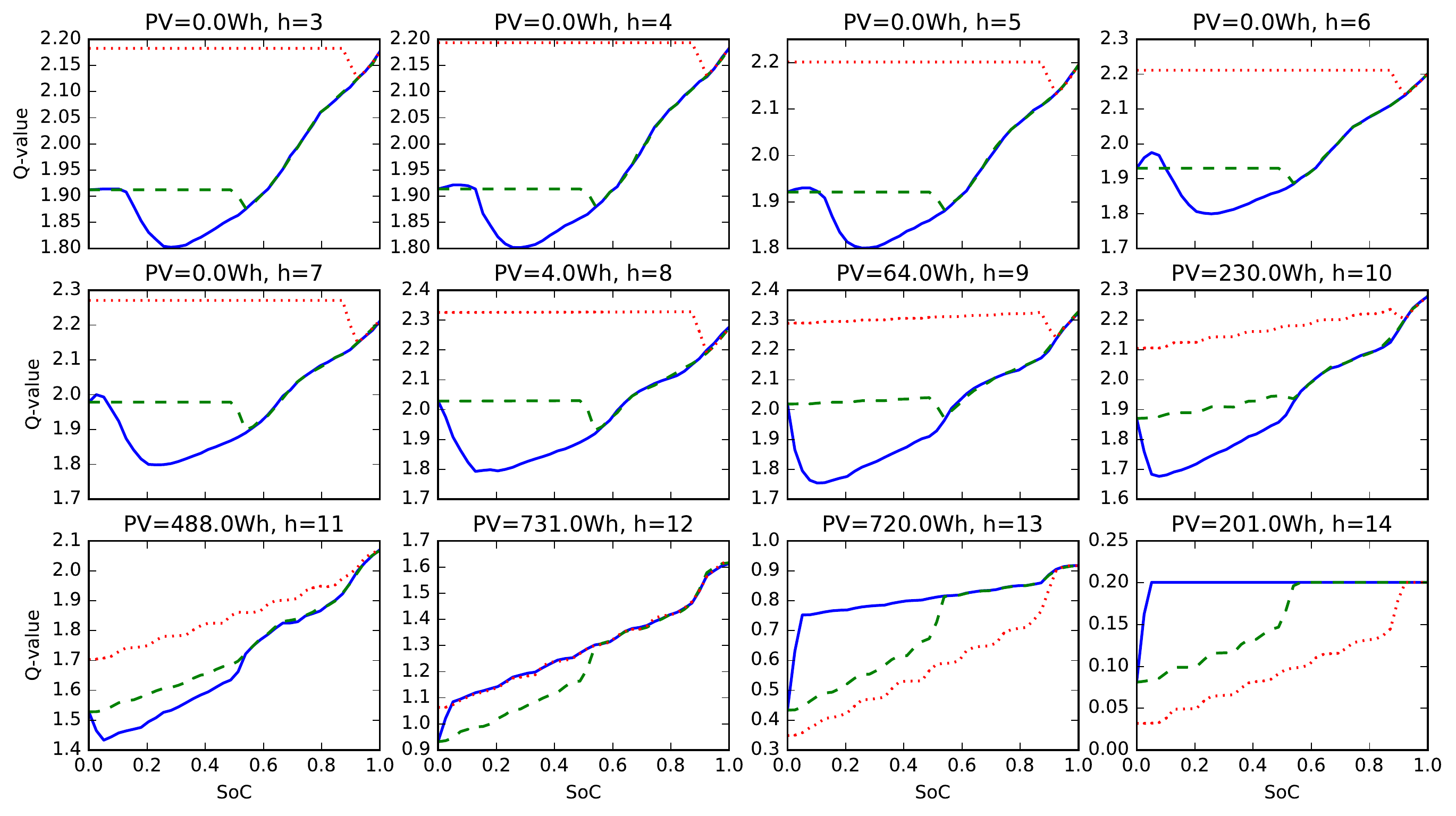}
\centering
\caption{Representation of the policy obtained by evaluating the Q-function over state-action space. Dotted line corresponds to heating to $T_{max}$ (u = 2). The dashed line corresponds to heating to $T_{min}$ (u = 1). Solid line corresponds to delay heating (u = 0). The predicted PV production in kWh and the hour of the day is indicated on top of every plot. For every SoC and time step the best action corresponds to the one with the minimum Q-value.}
\label{fig:Qmap}
\end{figure*}

A more in-depth analysis of the policy can be obtained by evaluating the $\hat{Q}$-functions over the state space for every action. Fig. \ref{fig:Qmap} presents the evaluated $\hat{Q}$-functions for the last 12 hours of the same day as in Fig. \ref{fig:policy}. It shows what the best SoC is at every time step and which action should be chosen.

\subsection{Performance Indicators}
To compare the PV self-consumption of the proposed control algorithm with the default thermostat control, performance indicators were calculated for both cases. Over a period stretching from September 2016 to January 2017 the scenarios were alternated each week. All the houses are controlled by the same algorithm. By switching the scenarios weekly, the seasonal influence of  meteorological conditions was minimized.

\begin{figure}
\centering
\includegraphics[width=0.48\textwidth]{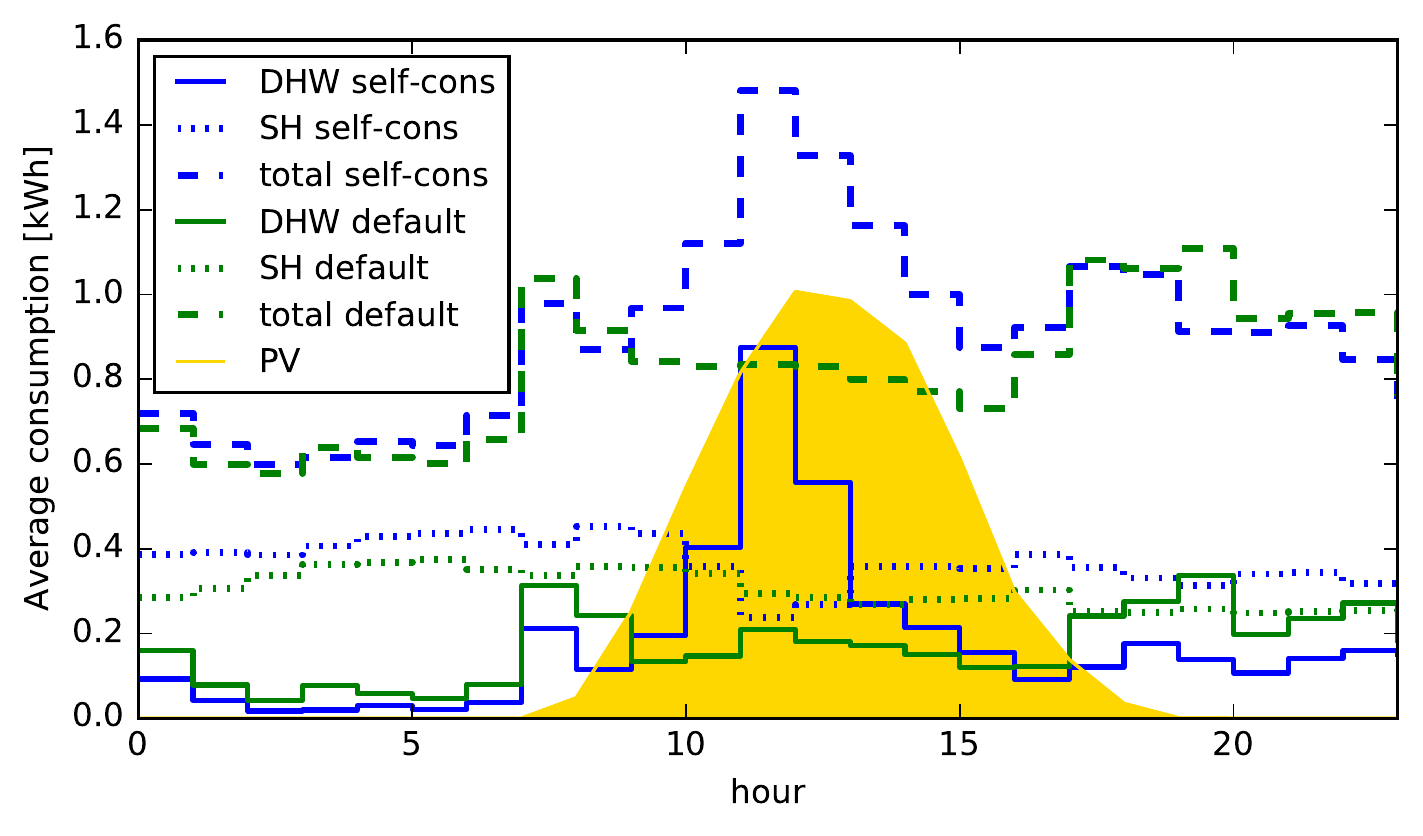}
\caption{Average DHW, SH and total consumption at every hour of the day for PV self-consumption (blue lines) and default thermostat period (green lines).}
\label{fig:average_loads}
\end{figure}

Table \ref{tab:perf_ind} shows some performance indicators for the six houses combined. It can be seen that the percentage of PV production captured by the heating of DHW almost triples by using the active control algorithm. The second indicator represents the share of PV production captured by the total electricity consumption. Non-controllable loads have been neglected in the control strategy since they are hard to predict for an individual house. However, in order to verify the practical importance of the algorithm, they are taken into account in PV captured by total. The increase in PV self-consumption of more than 20\%,  clearly demonstrates the effectiveness of the described approach. The four remaining indicators serve to compare certain important variables of the two scenarios. It can be observed that the electricity consumption is higher during the PV self-consumption test period, compared to when the thermostat was used. This difference is mainly caused by the increased space heating consumption. However, the table shows that the PV captured by space heating (SH) only increases with 1.1 percentage point. Taking into account this difference, there is still an increase of more than 20\% in total PV self-consumption. Fig. \ref{fig:average_loads} shows a qualitative representation of the results. It can be seen that the DHW consumption is clearly shifted towards the PV production.

As for now, the results only span a limited period of time (September-January). However, the experiment is still running and future results will provide a more elaborated and valid investigation over a longer time period covering all seasons.

\begin{table}
\caption{Performance Indicators.}
\label{tab:perf_ind}
\centering
\begin{tabular}{lrr}
\toprule
{} &  Thermostat & PV Self-Consumption \\
\midrule
PV captured by DHW:           &     6.25\% &  16.94\% \\
PV captured by total:         &    46.69\% &  58.52\% \\
PV captured by SH:            &    18.62\% &  19.71\% \\
El consumption per day:   &    20.17 kWh &  22.13 kWh\\
PV production per day:    &     5.61 kWh &   5.12 kWh\\
SH consumption per day:   &     7.38 kWh &   8.79 kWh\\
\bottomrule
\end{tabular}
\end{table}

\section{Conclusion}
\label{sec:conclusion}
This work has presented the implementation of a data-driven control approach for demand response in real-life residential buildings. The objective was to optimally schedule the heating cycles of the DHW buffer to maximize the self-consumption of the local PV production. A model-based reinforcement learning algorithm was deployed to tackle the underlying sequential decision making problem. The approach to learn stochastic occupant behavior, predict the PV production and take into account the dynamics of the system is discussed. A real-life experiment, using the proposed approach over a period of four months, with six residential buildings is analyzed. Each house is equipped with a smart heat pump and PV panels, dimensioned to achieve annual energy neutrality. The results show that the deployed algorithm significantly increases PV self-consumption compared to thermostat control.

In future work, more houses will be added to the experiment and the results will be evaluated over a longer period. Besides that, the influence of the assumptions made will be assessed by using different transition function models and it will be compared with a model-free approach.


\section*{Acknowledgment}
The research leading to these results has received funding from the
European Commission in the H2020 Programme - EU.2.1.5.3 under grant
agreement nr.680603 - REnnovates.



\bibliographystyle{IEEEtran}
\bibliography{references}
%
%
%
%
%

\end{document}